\documentclass{vgtc}   

\usepackage{mathptmx}
\usepackage{graphicx}
\usepackage{times}
\usepackage{amsmath}
\usepackage{algorithm}                 
\usepackage{algpseudocode}

\newenvironment{tight_enumerate}{\begin{enumerate} \itemsep
-2pt}{\end{enumerate}}

\usepackage[bookmarks,backref=true,linkcolor=black]{hyperref} 
\hypersetup{
  pdfauthor = {},
  pdftitle = {},
  pdfsubject = {},
  pdfkeywords = {},
  colorlinks=true,
  linkcolor= black,
  citecolor= black,
  pageanchor=true,
  urlcolor = black,
  plainpages = false,
  linktocpage
}

\onlineid{0}

\vgtccategory{Technique}

\vgtcinsertpkg

\title{Semantic Resizing of Charts Through Generalization:\\ A Case Study with Line Charts}

\author{Vidya Setlur\thanks{e-mail: vsetlur@tableau.com}\\ %
        \scriptsize Tableau Research %
\and Haeyong Chung\thanks{e-mail: haeyong.chung@uah.edu}\\ %
     \scriptsize University of Alabama in Huntsville %
}

\abstract{Inspired by cartographic generalization principles, we present a generalization technique for rendering line charts at different sizes, preserving the important semantics of the data at that display size. The algorithm automatically determines the generalization operators to be applied at that size based on spatial density, distance, and the semantic importance of the various visualization elements in the line chart. A qualitative evaluation of the prototype that implemented the algorithm indicates that the generalized line charts preserved the general data shape, while minimizing visual clutter. We identify future opportunities where generalization can be extended and applied to other chart types and visual analysis authoring tools.

} 


\CCScatlist{
  \CCScatTwelve{Human-centered computing}{Visualization}{}
}

   \teaser{
   \centering
    \includegraphics[width=8cm]{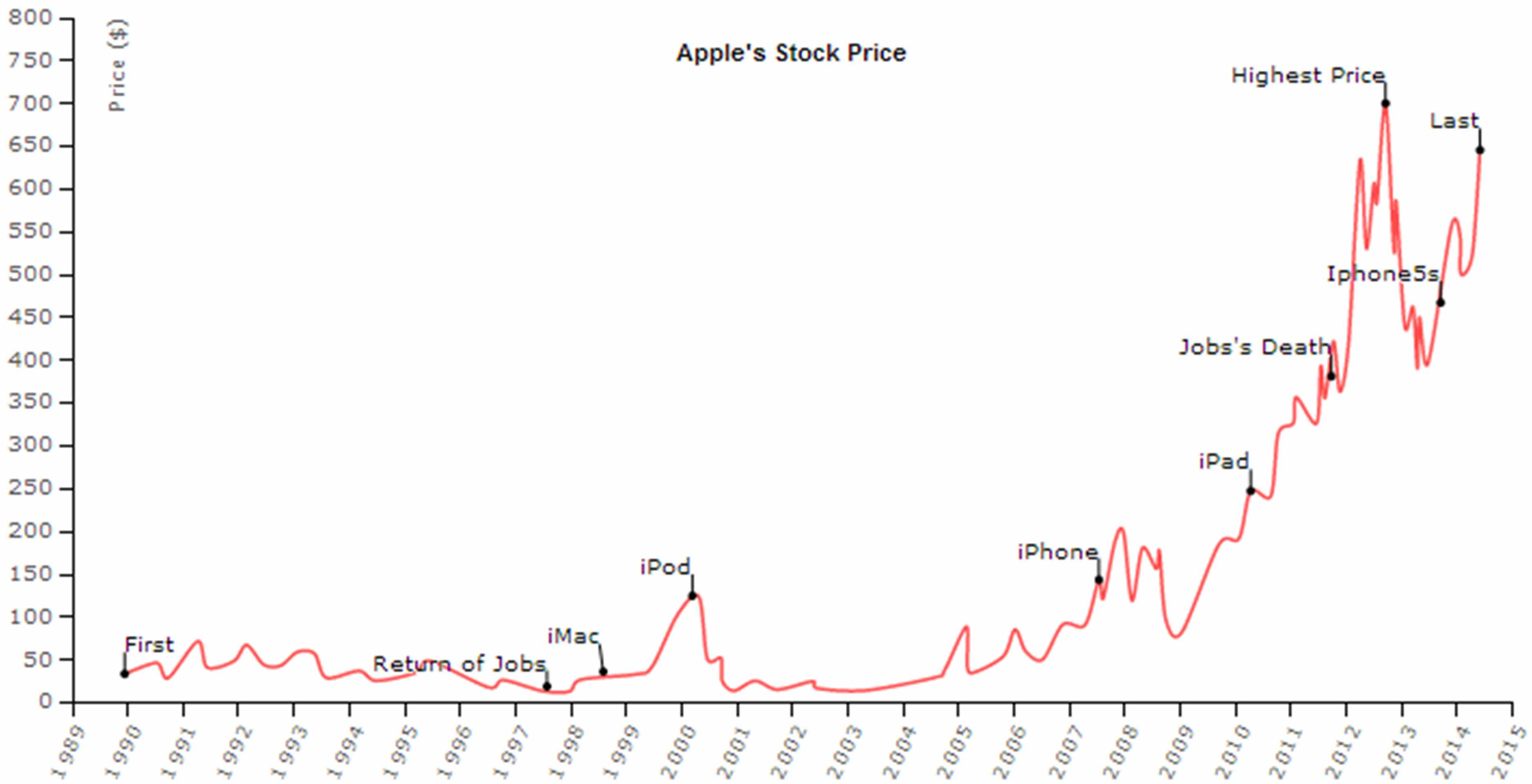}
    \hspace{1.5mm}
    \includegraphics[width=4.3cm]{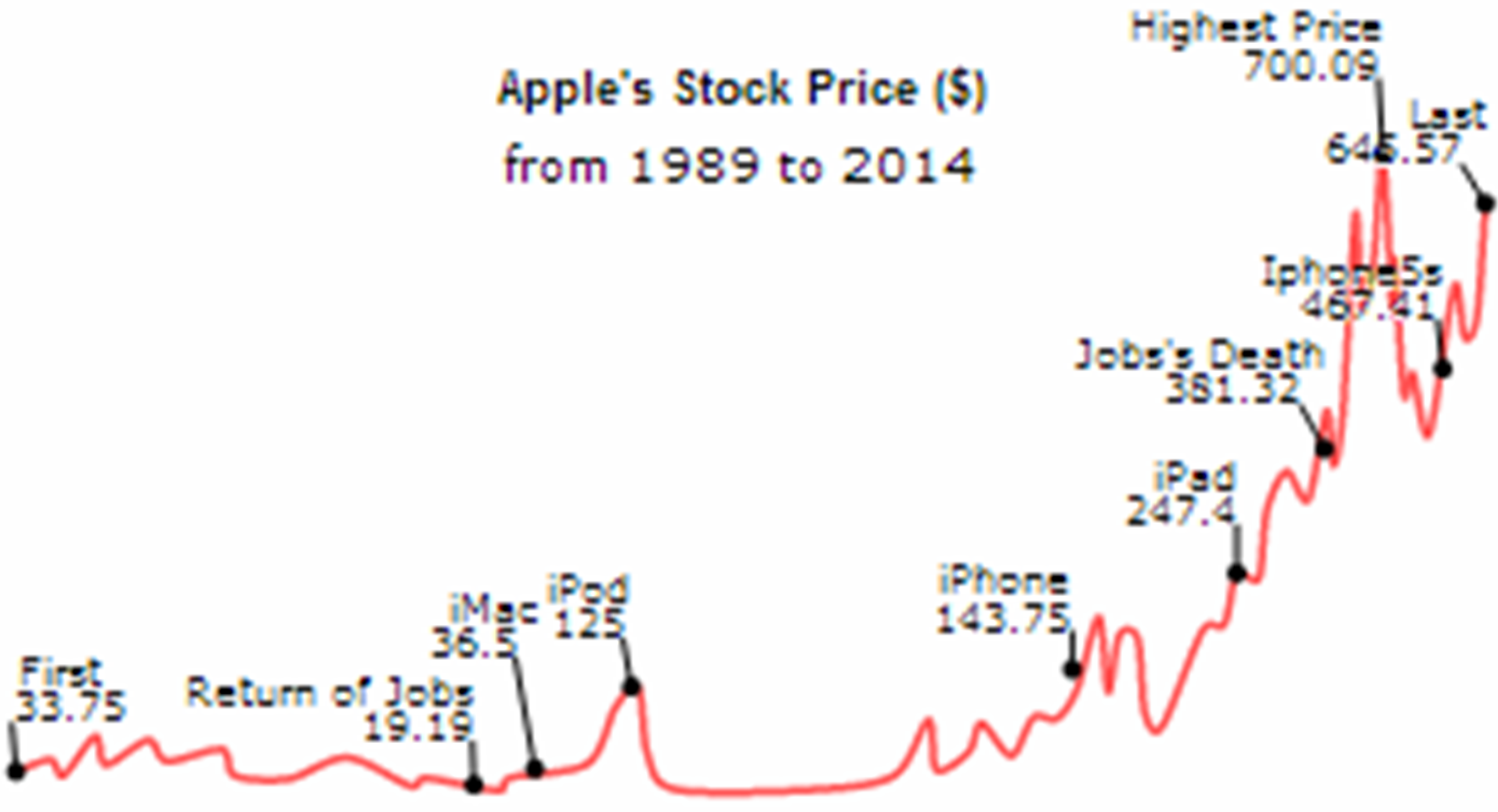}
    \hspace{1.5mm}
    \includegraphics[width=1.7cm]{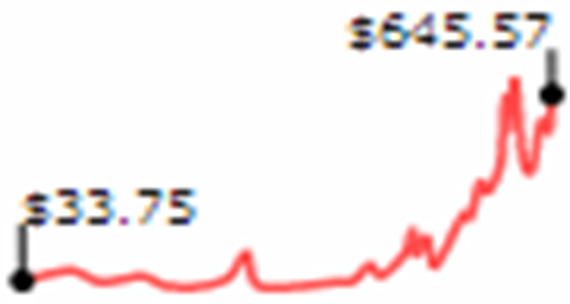}
   \caption{An example of a line chart automatically generalized to different display sizes. The algorithm preserves the various elements of the line chart based on their semantic importances at a given display size.}
   \label{fig:teaser}
  }

\begin{document}



\maketitle

\section{Introduction}
In the context of presenting data visualizations, resizing is particularly critical when engaging with a dashboard featuring limited screen real-estate, and/or when visualizations created on one display must then rendered on a different-sized display. The major challenge associated with developing techniques that facilitate resizing and creating multi-scale visualizations, is the significant number of variations that must be considered to represent a rescaled visualization effectively. A visualization author may need to represent every detail of the visualization at a given display size, while also taking into account every possible combination of the display properties such as resolution, size, and aspect ratio.

General resizing techniques such as uniform scaling and scale-and-stretch can be easily applied to resizing a visualization, but they tend to make the visualization illegible and increase the amount of visual clutter at smaller scales \cite{Wu}. Such resizing also does not consider semantic information represented by the underlying data. Hence, it is crucial for visualization techniques to support a smarter way to automatically adapt visual representations so that the viewer can read the chart more easily, regardless of the particular display size.

In cartographic literature, \emph{generalization} refers to the process of abstracting the visual detail in a map, with the goal of maintaining the legibility of the map at any given scale~\cite{Shea}. Inspired by cartographic generalization, this work explores how these principles can be applied to resizing data visualizations effectively at different scales. Particularly, we extend cartographic generalization to rescaling line charts, a common chart type. We present a set of spatial metrics to examine geometric properties and relationships among elements in a line chart. These  metrics are used to determine the presence of visual clutter and complexity in a view. Based on these metrics, we developed a set of generalization techniques for semantically resizing line charts to target display sizes.

\section{Related Work}

There are several approaches to resizing content beyond mere uniform scaling and can be organized into the following categories:

\subsection{Cartographic Generalization}
Cartographic generalization enables one to simplify or eliminate less semantically important features, exaggerate more important ones, and resolve visual clutter to improve information quality on a smaller scale \cite{Brassel,McMaster,Shea}. Route maps have effectively and succinctly applied various forms of generalization \cite{Agrawala,zipf2002user}. Kray et al. describe a method of presenting route instructions on a mobile device depending on various situational factors such as limited resources and varying quality of positional information \cite{604066}.  Although automatic map generalization techniques have been described for two decades in the cartographic and geographic literature, little research has been undertaken on how these methods could be extended to other forms of visualization.

\subsection{Scroll, Pan and Zoom, Overview + Detail}
Scroll and pan behavior maintains the layout and size of the original viewport, allowing for a portion of the visualization to be seen at any given time~\cite{tableau,powerbi}, resulting in a loss of context \cite{Ball}. Semantic zoom lets the user see different amounts of detail based on semantic importance~\cite{cockburn}. Panning and zooming are often combined to create a continuous navigation experience~\cite{Bederson:1994}. Overview + detail techniques display more than one level of detail such that manipulating the overview causes a corresponding change in the detail view. In SeeSoft \cite{Eick:1992}, a miniaturized overview of text operates as a scrollbar for a more detailed view. While overview + detail reduce disorientation by having the overview co-present with the detail view, the techniques often suffer from visual discontinuity between the two views. Schwab et al.~\cite{Schwab:2019} studied how pan and zoom timelines are used to navigate large time series data. They found that the visual context and orientation play a role in efficient navigation. While these interaction techniques provide users the flexibility of viewing a portion of the content at a time, the navigation can be often disorienting especially when the frame of reference is not clearly defined. Our work explores how line charts can be generalized for a specific target size, displaying detail for more salient information.

\subsection{Content Retargeting}
For viewing large-sized content on smaller screens, there has been work on adapting or \textit{retargeting} the content for these smaller displays. Automatic reformatting of web pages concatenates columns or hides less important information \cite{Chen:2003, Trevor:2001}. With the prevalance of online data journalism, Kim et al.~\cite{kim:2019} explored how data thumbnails in these articles can be designed to be interpretable. MiniMap \cite{minimap} changes the size of the text relative to the size of the viewport. Other techniques explore changing the geometrical properties of visualization elements and views, including geometrical transformation \cite{Avidan,Wang,Wolf} and view deformation \cite{Wu,Baldassi}.  Vistribute~\cite{vistribute} automatically distributes visualizations and user interface components across multiple heterogeneous devices. Shi et al.~\cite{shi:2006evaluation} evaluate a diversity of automatic line simplification algorithms based on positional accuracy and processing time. These various techniques for retargeting however, may get challenging if the layout transformations affect the semantics of the content. We explore generalization as a way to preserve the recognizability of semantically important elements in the line chart at different display sizes.

\subsection{Responsive Visualizations}
The prevalence of mobile devices motivates the need to design communicative visualizations that are responsive to varying screen sizes. Kim et al.~\cite{kim2021design} identify strategies and trade-offs between the presentation of information and the intended takeaway of the chart. Often these visual representations and graphical user interfaces have to be adapted for smaller displays and lower resolution \cite{blascheck2018glanceable, drucker2013touchviz, brehmer2018visualizing}. ViSizer~\cite{Wu} presents a perception-based framework for scaling important regions uniformly and deforming homogeneous context. MobileVisFixer~\cite{wu2020mobilevisfixer} adapts a reinforcement learning-based approach that automatically learns and applies decision rules for generating mobile-friendly visualizations. Hoffswell et al.~\cite{hoffswell2020} analyzed a corpus of responsive news visualizations that informed a prototype tool for designing responsive visualizations for different device contexts. Our work is motivated by similar goals, but we specifically explore how cartographic generalization principles can be applied to line charts. By considering the spatial relationships of the individual elements in the chart, we developed an automated approach to emphasize and de-emphasize information based on their semantics and target display size.


\section{Algorithm}
Our algorithm identifies and categorizes chart elements in a given view and adaptively performs a set of generalizations that selects and abstracts the elements based on a set of spatial and semantic constraints. The constraints determine the type of generalization operation that is applied to the resized line chart to minimize visual clutter. We implemented our algorithm using HTML5/JavaScript and D3~\cite{bostock2011d3}.  We now discuss the algorithm in more detail.

\subsection{Identifying Semantic Importance}
The first step in our algorithm is to identify all of the elements in the given chart. The algorithm assigns higher semantic weights to local extrema, first and last data values compared to chart elements such as axes or tick marks. Each element is assigned a unique ID and the algorithm computes a bounding box for each of them. Similar elements are then categorized into layers. This allows us to apply different constraints and generalization operators based upon the unique characteristics and semantics of each element layer. Each layer is assigned a value from 0 (lowest importance) to 1 (highest importance) based on the semantics of a line chart~\cite{Kosslyn}.

\subsection{Compute Spatial Metrics}
The algorithm then computes various spatial metrics for the elements in the resized line chart to minimize visual clutter. Based on visualization best practices~\cite{MacEachren,jockbook,Tufte}, the amount of visual clutter is informed by the following guidelines:

\noindent \textbf{Avoid congestion:} A visualization view should not include too many elements at specific region. 

\noindent \textbf{Avoid conflict:} To maintain legibility, elements should not overlap. Each element should be easily identifiable and readable. 

\noindent \textbf{Make more semantically important elements prominent:} An element of higher importance value should be more visible than less important ones. 

\noindent We now describe each of these spatial metrics in more detail.

\subsubsection{Density}

\begin{figure}[ht]
\centering
\includegraphics[width=3.4in]{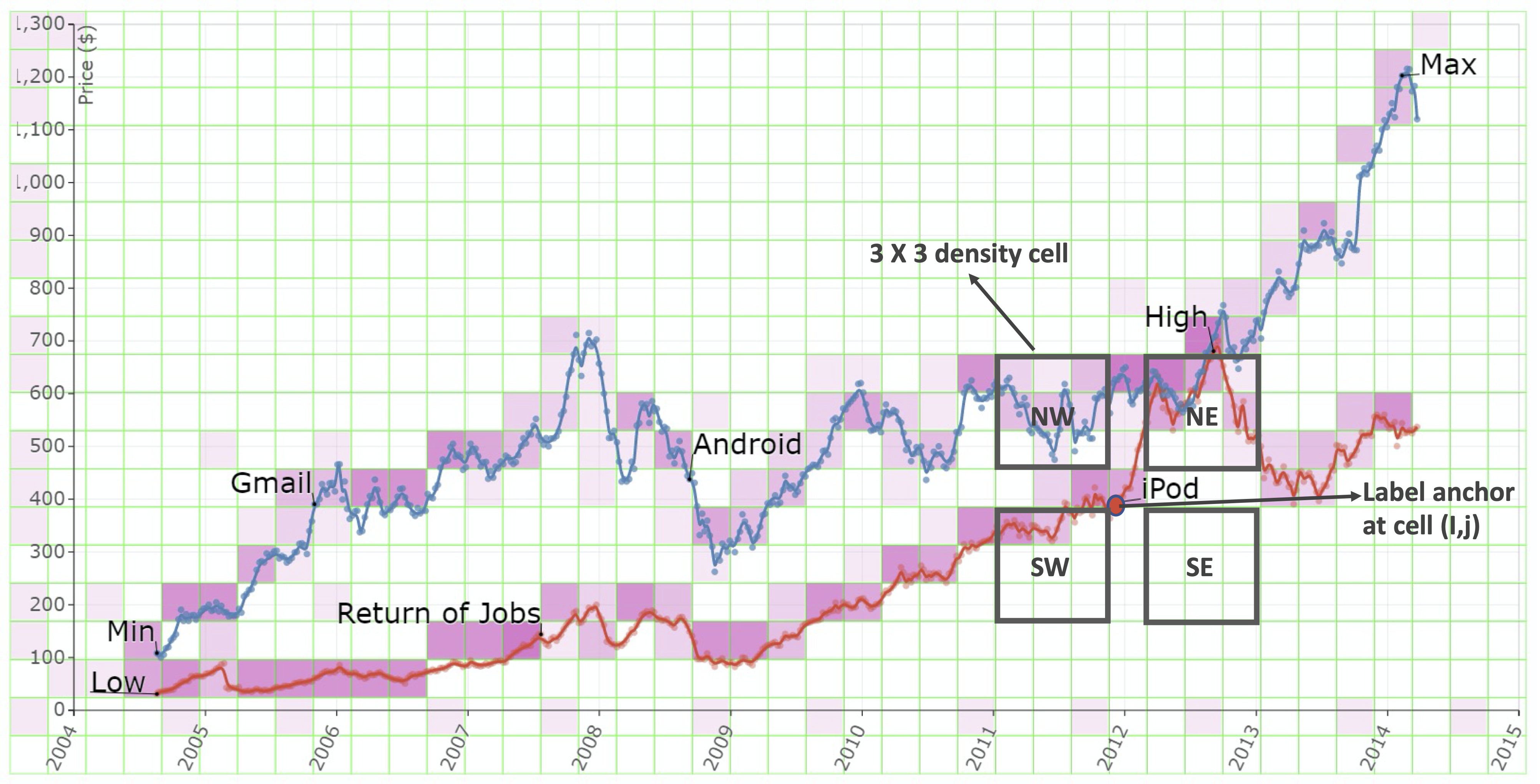}
\caption{Computing density. The entire multivariate line chart view is divided into $n \times m$ uniform cells and density at each cell is calculated to determine the number of elements visible within each cell area. Darker the square, higher the density.} 
\label{fig:density}
\end{figure}
Data density can provide simple, yet effective metrics for evaluating visual clutter in the view~\cite{Tufte, Rosenholtz}. To calculate density, we divide the entire visualization view into uniform $n \times m$ cells (where $m$ and $n$ are empirically determined based on the target display size). We then apply Topfer's Radical Law \cite{Topfer,Woodruff} to express the number of elements that can be maintained at that size (Figure \ref{fig:density}):
\vspace{-1mm}
\begin{equation}
\mbox{Cell information density}  =\frac{\mbox{Number of Elements}}{\mbox{Number of Pixels in the Cell}}
\end{equation}


This metric allows for generalization operators to be applied to different regions based on their density value.  For example, we can show more information or enlarge elements in less dense regions.

\subsubsection{Distance}
In our algorithm, the distance is measured between similar elements within the same layer such as labels or tick marks. The distance is measured using Euclidean distance.  We utilize this metric to assess whether elements (e.g., annotations and associated data points) are too close to each other.

\subsubsection{Collision}
This metric determines collisions between elements in a chart that contribute to visual clutter. We employ a quadtree, a compact data structure to keep the algorithm performant~\cite{klinger1971patterns}. We use the following equation to compute the overlapping area between elements:
\begin{equation}
\mbox{Collision area} = \sum_{i}^{N}\sum_{j \neq i}^{N}A(i,j)
\end{equation}

where $A(i,j)$ is the area of overlap between elements $i$ and $j$. 

\subsubsection{Area ratio}
 As the display size decreases, the proportion of areas that elements occupy with respect to the area of the resized chart, increases. This metric assesses the ratio of the total area of elements to the area of the entire visualization view and is used to maintain the area of the more semantically important elements in the chart, deemphasizing the area of less important ones. Area ratio is computed as follows:
 
\begin{equation}
\mbox{Area ratio} =  \frac{\mbox{Area of element}}{\mbox{Total display area}}
\end{equation}


\subsection{Generalization operators}
The primary goal of generalization is to maintain the recognizability of important elements, while deemphasizing less important information. We describe four generalization operations based on the spatial metrics - jittering, elimination, simplification, and merging. 

\subsubsection{Jittering}
Jittering is a technique used to resolve collisions between elements by displacing the elements from their original positions to reduce visual clutter. We employ and extend the label placement simulated annealing algorithm~\cite{Wang} by adding additional heuristics:

\begin{tight_enumerate}
\item A label should not overlap a data point.
\item A label should be located at the place which has the lowest information density.
\item A label's text-anchor should be updated based on the new position of the label.
\end{tight_enumerate}

We first determine a $3 \times 3$ density cell diagonally adjacent to a label anchor located in cell $(i, j)$ with density (Figure \ref{fig:density}). We chose a $3 \times 3$ cell neighborhood as it provided a reasonable heuristic for determining visual clutter. The sum of each $3 \times 3$ density cells is computed as follows:
\vspace{-1mm}
\begin{equation}
\begin{split}
\mbox{Northwest (NW)} = d(i-3 .. i-1, j-3 .. j-1)  \\
\mbox{Northeast (NE)} = d(i+1 .. i+3, j-3 .. j-1)  \\
\mbox{Southwest (SW)} = d(i-3 .. i-1, j+1 .. j+3) \\
\mbox{Southeast (SE)} = d(i+1 .. i+3, j+1 .. j+3) \\
\end{split}
\end{equation}

where $d(a .. b, c .. d)$ denotes the sum of cell information density $d(x,y)$ for all $a \le x \le b$ and $c \le y \le d$. $i$ and $j$ are the indices of cells where the anchor is located and the screen coordinates where the origin is, at the top-left. After calculating the sum of each of the four directions, jittering is applied by calculating a new position that has the minimum density sum: $min (\mbox{NW}, \mbox{NE}, \mbox{SW}, \mbox{SE})$

\subsubsection{Elimination}

\begin{figure}[ht]
\centering
\includegraphics[width=2.0in]{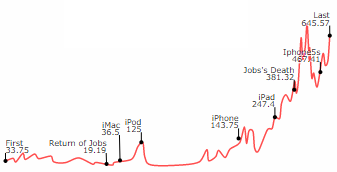}
\includegraphics[width=1.3in]{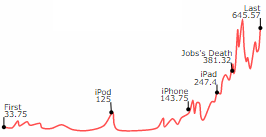}
\caption{Left: Before elimination. Right: After elimination. Here, a few intermediate labels are removed as they are less important than other labels denoting start and end points as well as local maxima.}
\label{fig:elimination}
\end{figure}

There is a high likelihood that too many elements will occupy a small area when the scale is reduced. This situation significantly increases both local information density and the likelihood of conflict. We compute a score $S$ based on semantic importance for each element to determine which of the elements ought to be removed.  

\vspace{-3mm}
\begin{equation}
\mbox{S} =  (1 - \mbox{imp}) \times W_{imp}  +\\
    \mbox {local density} \times  W_{dens} + overlap \times W_{ov}
\end{equation}

where $W_{imp}$, $W_{dens}$, and $W_{ov}$ are weights for the importance $imp$, local density, and overlap constraints respectively. Algorithm \ref{algo1} performs the jittering and elimination operations and Figure \ref{fig:elimination} shows how elimination is applied to remove less important labels.

\begin{algorithm}
	\caption{Elimination operation}
	\hspace*{\algorithmicindent} \textbf{Input:} {Resized line chart without elimination applied and semantic importance score $S$ for all chart elements.}\\
	\hspace*{\algorithmicindent} \textbf{Output:} {Resized line chart with elimination applied.}\\
	\begin{algorithmic}[1]
	  \While {! satisfyConstraints}
        \State Perform jittering to ascertain if constraints are satisfied. 
        \State Eliminate elements starting with lowest $S$ that overlap the most with other elements in the chart.
        \EndWhile
	\end{algorithmic}
	\label{algo1}
\end{algorithm}

\subsubsection{Simplification}

To reduce this complexity and clarify semantic intent, we simplify the line using the Douglas-Peucker algorithm~\cite{Douglas} line simplification method. We found that this line simplification method offers efficient compression ratios while retaining important visual features in the lines. As shown in Figure \ref{fig:teaser}, simplification maintains the overall data shape, while preserving visually prominent features.

\subsubsection{Merging}
\begin{figure}[ht]
\centering
\includegraphics[width=1.7in]{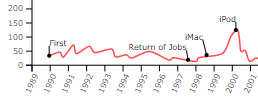}
\includegraphics[width=1.6in]{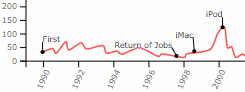}
\caption{Left: Before merging. Right: After merging. Here, the domain of the x-axis is updated to merge yearly tick marks to two-year intervals.}
\label{fig:combine}
\end{figure}

This operation combines elements when there is congestion or the elements are in conflict. In the context of line charts, this operation applies to tick labels. As shown in Figure \ref{fig:combine}, two tick labels are merged into a single tick label by updating the axis domain and the interval between the tick marks.  

\subsection{Performance and space complexity}
With our generalization algorithm, a new chart can be rendered on the fly based on the spatial constraints of the current view. While the overall complexity of the algorithm depends on the actual generalization operators employed, the complexity is at most $O(n^2)$ based on the worst-case time-complexity of the jittering algorithm. The algorithm has been tested on a desktop browser for various target sizes and further investigation would need to be done to assess the algorithm's performance on mobile devices and smart watches.

\section{Evaluation}
We conducted an evaluation of the algorithm with the following goals: (1) collect qualitative feedback on the usefulness of the generalization algorithm on various target display sizes and (2) identify limitations and future opportunities. Because the main goal of our study was to gain qualitative insight in the algorithm behavior, we encouraged participants to think aloud with the experimenter.

\subsection{Method}
\subsubsection{Participants}
We recruited $30$ volunteers ($12$ males, $18$ females, age 24 – 54) from a local town mailing list. The participants had a variety of backgrounds  - software engineer, technical writer, sales consultant, product manager, program manager, and graduate student. Based on self-reporting, all were fluent in English and had experience reading line charts. $22$ used a visualization tool~\cite{tableau,powerbi} on a regular basis and the rest considered themselves having limited proficiency.

\subsubsection{Procedure and Apparatus}
We ran a between-subjects design study where each participant was randomly assigned one resized chart from a set of $10$ single-line charts. We sourced the charts from Pew Research~\cite{pewresearch}, Wikipedia~\cite{wikipedia}, and Tableau Public~\cite{tableaupublic} and recreated them in D3. We removed graphical elements that could potentially affect the readability of the features such as highlighting and background shading. All the original charts had a display size of $6307 \times 3220$. \\
\noindent \textbf{Task 1:} During the first part of the study, each participant was initially shown the original line chart and then randomly shown one of three resized images of the chart, targeted at tablet ($1536 \times 2048$), phone ($750 \times 1334$), and watch ($324 \times 394$) display sizes. The display sizes were chosen based on evaluation criteria commonly used to evaluate visualizations on mobile devices~\cite{blumenstein2016}. Because we were seeking subjective responses, participants could complete only one trial to avoid biases that might arise from repeated exposure to the task. To assess whether features in the chart were discernible, participants were asked the values for the start and end, local extrema, minimum, and maximum data points. We did not impose a constraint on the amount of time spent looking at the chart and answering the questions, emulating chart reading in the real world. 

\noindent \textbf{Task 2}: The second part of the study was exploratory where participants used the prototype with the original line chart loaded into a web browser. They could resize the chart to any arbitrary display size by dragging and resizing the browser window. The study concluded with an interview. 

The prototype was hosted on the experimenter's 16-inch laptop, a $2.4$ GHz MacBook Pro running macOS Catalina $10.15.7$ set to a resolution of $3072 \times 1920$. Each session took about $45$ minutes.

\subsubsection{Analysis Approach}
 We employed a mixed-methods approach involving qualitative and quantitative analysis, but considered the quantitative analysis mainly as a complement to our qualitative findings.

\section{Results and Discussion}
Overall, participants were positive about the system and identified many benefits for the system being able to dynamically generalize the line chart to different display sizes. Several participants were impressed with the system's ability to preserve local extrema features and the general data shape in the line chart at smaller display sizes -  (``I can recognize the smaller chart and has kept the peaks that I observed in the larger one ($P2$).''). All the participants were able to identify the start and end values in the three generalized line charts during Task 1. $100\%$ for the larger ($1536 \times 2048$) size, $80\%$ (8/10) for the phone (750×1334) size, and $30\%$ (3/10) of the participants for the watch ($324\times 394$) size were able to accurately provide values for the local extrema, minium, and maximum points. Participants appreciated the sparkline appearance in the smallest chart and commented that such a representation would be useful for certain applications. For example, ``I go running and this chart can be useful to show my heart rate on my iWatch. ($P25$)'' and ``I see these charts a lot for stock prices. They are simple and easy to read at a quick glance ($P14$).'' During the exploratory part of the study (Task 2), participants appreciated the dynamic nature of the algorithm to be able to generalize the chart as they resized the browser window. Comments related to this behavior included, ``I can see this being very useful if I'm switching between devices and want to view the same data ($P2$)'' and ``I often need to add charts into my table. It would be neat if I can drag a line chart into a cell and it creates a little microchart for me ($P10$).''

The study has limitations and provides opportunities for further improving and expanding the scope of the generalization behavior.\\
\noindent \textbf{Exploring complex chart types and tasks}: We evaluated our generalization algorithm for univariate line charts. While the spatial metrics apply to multivariate line charts, there are interesting research opportunities to explore how well generalization applies to other chart types. $P4$ said, ``I use a lot of Excel tables, but often struggle to get the gist of them when I view them on my mobile device.'' Exploring how data distributions can affect the spatial metrics and semantic properties of chart elements, is another interesting direction of research inquiry. The task of reading data values from the charts at different scales, is rather simplistic. Future work should explore more complex chart reading tasks including trends, seasonal patterns, and overall takeaways.\\
\noindent\textbf{Control over feature retention and simplification}: A challenge for automating the generalization process is representing data at multiple scales~\cite{buttenfield}. Participants wanted to have more control over what information is preserved as the line chart is resized. $P7$ commented, ``I wanted to keep the label on the small dip in the dollar price as I felt that was important and I wish there was a way to keep it sticky.'' Similar to responsive web design tools~\cite{responsivewebdesign} that support both automation and flexibility of content and layout, there is a need to explore responsive chart and dashboard authoring tools.\\
\noindent \textbf{Generalization with deep learning models}: Our approach employs spatial metrics for determining the generalization operation. Neural networks and deep learning methods are promising avenues for learning good generalization examples through pattern recognition~\cite{Karsznia:2020}. Exploring how such deep learning approaches can replicate many difficult aspects of expert generalization, such as line smoothing, enlargement, and displacement, would help scale the applicability of chart generalization over a variety of use cases. 

Finally, future work should test these ideas in a live system for specific real-world analytical tasks and actual target display devices rather than in the artificial setting of a study.

\section{Conclusion}
This  paper  presents  a  technique  for applying cartographic principles to the resizing of line charts at different target display sizes. Specifically, we present a set of spatial metrics to examine geometric properties and relationships among elements in line charts. We employ these metrics to determine a set of generalization operations for semantically resizing line charts; maintaining the legibility of visually prominent features and demphasizing less important features. An evaluation of the algorithm indicates that participants found that the generalized line charts preserved the general data shape and features from that of the original. Feedback from interacting with the prototype identified opportunities for further exploring the space of generalization techniques for creating responsive chart design. 

\newpage
\bibliographystyle{abbrv}
\bibliography{references}
\end{document}